\documentclass[aps,prc,twocolumn,superscriptaddress,nofootinbib,
preprintnumbers,showkeys]{revtex4-2}

\usepackage[utf8]{inputenc}
\usepackage[english]{babel}
\usepackage{amssymb,amsthm,amsmath,amstext,amsbsy,amsopn,mathrsfs}
\usepackage{bbm}
\usepackage{nicefrac}
\usepackage{slashed}
\usepackage{graphicx}
\usepackage{hyperref}
\usepackage{leftidx}
\usepackage{environ}
\usepackage{mathtools}
\usepackage{xspace}
\usepackage{array}
\usepackage{xcolor}
\usepackage{pgfplotstable}
\usepackage{isotope}
\usepackage{suffix}
\usepackage{booktabs,colortbl}
\usepackage{physics}
\usepackage{enumitem}
\usepackage{aligned-overset}

\newcommand{\ie}{\textit{i.e.}\xspace}

\newcommand{\apriori}{\textit{a priori}\xspace}
\WithSuffix\newcommand\apriori*{\textit{a-priori}\xspace}

\newcommand{\mathspace}{\ \ }
\newcommand{\mathtext}[1]{\mathspace\text{#1}\mathspace}

\newcommand{\vecr}{\mathbf{r}}
\newcommand{\vecx}{\mathbf{x}}
\newcommand{\vecy}{\mathbf{y}}
\newcommand{\vecz}{\mathbf{z}}
\newcommand{\vecn}{\mathbf{n}}
\newcommand{\vecm}{\mathbf{m}}

\newcommand{\vecq}{\mathbf{q}}

\newcommand{\vecR}{\mathbf{R}}

\newcommand{\hatx}{\hat{\vecx}}
\newcommand{\haty}{\hat{\vecy}}
\newcommand{\hatz}{\hat{\vecz}}

\renewcommand{\dv}[2]{\frac{\dd #1}{\dd #2}}

\newcommand{\ddr}{\mathrm{d}^3\vecr}

\newcommand{\ii}{\mathrm{i}}
\newcommand{\ee}{\mathrm{e}}

\newcommand{\ZZ}{\mathbb{Z}}
\newcommand{\RR}{\mathbb{R}}

\newcommand{\Rp}{\mathrm{Re}}

\newcommand{\Ei}{\text{Ei}}

\newcommand{\mean}[1]{\langle #1\rangle}

\newcommand{\matrixelem}[3]{\left\langle #1 \middle\lvert #2 \middle\lvert #3 \right\rangle}
\WithSuffix\newcommand{\matrixelem}*[3]{\big\langle #1 \big\lvert #2 \big\lvert #3 \big\rangle}

\newcommand{\Trans}{\hat{T}}
\newcommand{\Rot}{\hat{R}}

\newcommand{\setball}{B}
\newcommand{\setcube}{C}
\newcommand{\setasymptotic}{A}

\newcommand{\Einf}{{-}E_\infty}
\newcommand{\Einfp}{E_\infty}

\newcommand{\EL}{-E(L)}

\newcommand{\Eshift}{\Delta E(L)}

\newcommand{\psiinf}{\psi_\infty}
\newcommand{\psiinfa}{\psi_{\infty,\text{asm}}}

\newcommand{\psiEasm}{\psi_{E,\text{asm}}}
\newcommand{\psiL}{\psi_L}
\newcommand{\psiLasm}{\psi_{L,\text{asm}}}
\newcommand{\psiLint}{\psi_{L,\text{int}}}
\newcommand{\psiasm}{\psi_\text{asm}}
\newcommand{\psiint}{\psi_\text{int}}

\newcommand*\rvec[1]%
{\ensuremath{\overset{\smash{\raisebox{-1.5pt}{\tiny$\rightarrow$}}}{#1}}}
\newcommand*\lvec[1]%
{\ensuremath{\overset{\smash{\raisebox{-1.5pt}{\tiny$\leftarrow$}}}{#1}}}

\NewEnviron{subalign}[1][]{%
\begin{subequations}\begin{align}
  \BODY
\end{align}\label{#1}\end{subequations}
}

\NewEnviron{spliteq}{%
\begin{equation}\begin{split}
  \BODY
\end{split}\end{equation}
}

\NewEnviron{mleq}{%
\begin{equation}
  \BODY
\end{equation}
}

\newcommand{\higherorder}{\order{e^{{-}\sqrt{2}\kappa L}}}

\begin{document}

\title{Radius extrapolations for two-body bound states in finite volume}

\author{Anderson Taurence}
\email{ajtauren@ncsu.edu}
\affiliation{Department of Physics, North Carolina State University,
Raleigh, NC 27695, USA}

\author{Sebastian König}
\email{skoenig@ncsu.edu}
\affiliation{Department of Physics, North Carolina State University,
Raleigh, NC 27695, USA}

\begin{abstract}
 Simulations of quantum systems in finite volume have proven to be a useful tool
 for calculating physical observables.
 Such studies to date have focused primarily on understanding the volume
 dependence of binding energies, from which it is possible to extract
 asymptotic properties of the corresponding bound state, as well as on
 extracting scattering information.
 For bound states, all properties depend on the size of the finite
 volume, and for precision studies it is important to understand such effects.
 In this work, we therefore derive the volume dependence of the mean squared
 radius of a two-body bound state, using a technique that can be generalized
 to other static properties in the future.
 We test our results with explicit numerical examples and demonstrate that we
 can robustly extract infinite-volume radii from finite-volume simulations
 in cubic boxes with periodic boundary conditions.
\end{abstract}

\maketitle

\section{Introduction}
\label{sec:Introduction}

Finite-volume (FV) simulations of quantum systems in periodic cubic boxes are
a powerful tool that is used to study properties of nuclear bound states and
scattering.
A series of highly influential
papers~\cite{Luscher:1985dn,Luscher:1986pf,Luscher:1990ux} established in the
1980s and 90s that real-world properties of a quantum system are encoded in how
its discrete energy levels change as volume size is varied.
Over the past decades, this fruitful idea has spurred a lot of activity, with
recent focus on the study of three-body systems~\cite{Kreuzer:2010ti,Kreuzer:2012sr,Polejaeva:2012ut,Briceno:2012rv,%
Kreuzer:2013oya,Meissner:2014dea,Hansen:2015zga,Hammer:2017uqm,%
Hammer:2017kms,Mai:2017bge,Doring:2018xxx,Pang:2019dfe,Culver:2019vvu,%
Briceno:2019muc,Romero-Lopez:2019qrt,Hansen:2020zhy,Muller:2021uur,%
Draper:2023xvu,Bubna:2023oxo}, motivated primarily by applications to Lattice
Quantum Chromodynamics (Lattice QCD).
For two-cluster bound states, the volume dependence of the binding energy is
known for an arbitrary number of constituents~\cite{Konig:2017krd}.
Recent work derived the volume dependence of two-body bound states comprised of
charged particles, with full nonperturbative account for the repulsive Coulomb
interaction~\cite{Yu:2022nzm}, and Ref.~\cite{Yu:2023ucq} extended the
method to resonances.
An important motivation for understanding the volume dependence of bound
states is that knowledge of the functional form makes it
possible to extract asymptotic normalization coefficients (ANCs) from FV
calculations, for example based on Lattice Effective Field Theory (Lattice EFT)
simulations of atomic nuclei~\cite{Lee:2008fa,Lahde:2019npb,%
Lu:2021tab,Shen:2022bak,Elhatisari:2022qfr}.

In this paper, we extend studies of the volume dependence for bound states
beyond what is known for binding energies.
As simplest observable, we consider mean squared radii, $\mean{r^2}$, of
two-body bound states, defined (in more detail in the following section) as the
expectation
value of an operator that measures the average distance of the constituents
from their common center of mass.
Just like the binding energy, $\mean{r^2}$ will be shifted from its physical
value in FV, and the magnitude of this shift can be traced back to changes in
the wave function induced by being confined to a periodic box.
We derive in detail the functional form of the radius finite-volume shift, which makes it
possible to perform extrapolations from a set of finite-volume calculations to
the real world, \ie, infinite volume.

Unlike the binding energy, which is known to depend, to leading exponential
order, only on asymptotic properties of the wave function (and is thus
universal with respect to the details of the short-range interaction that
gives rise to the bound state~\cite{Luscher:1985dn,Konig:2011nz,%
Konig:2011ti,Konig:2017krd,Yu:2022nzm}), one should expect $\mean{r^2}$ to
be sensitive in principle to the form of the wave function at all relative
distances.
This has indeed been observed for radii and other static properties of
bounds states calculated in truncated harmonic oscillator
bases~\cite{Furnstahl:2012qg,Furnstahl:2013vda,Odell:2015xlw}.

We derive in this work analytical expressions for the leading volume
dependence of the mean-squared radius for two-body states bound by a
short-range interaction.
Based on an appropriate ansatz for the relevant volume dependence of the
wave function, we obtain explicit formulas for states within the $A_1^+$
and $T_1^-$ representations of the cubic symmetry group, which correspond
approximately to S- and P-wave states in infinite volume.
A constructive prescription is given for the general case.
Our results are relevant for example for Lattice QCD studies of the deuteron
radius, and the formalism we develop paves the way for deriving analogous
relations for other static observables, as well as for bound states
comprised of more than two particles.
Relations of this form will have applications not only in Lattice QCD
studies of multi-nucleon bound states, but also to precision studies of
atomic nuclei with Lattice EFT.

This paper is organized as follows.
In Sec.~\ref{sec:Formalism} we develop the formalism for deriving the
finite-volume radius shift $\Delta\!\expval{r^2}(L)$, starting with a
discussion of how the bound-state wave function changes when it is confined
to the periodic box.
Subsequently, in Sec.~\ref{sec:Results} we present closed-form analytical
expressions that describe the radius shift for bound states within the
$A_1^+$ and $T_1^-$ cubic representations and we
verify these the results with
explicit numerical calculations.
Finally, in Sec.~\ref{sec:Conclusion} we close with a summary and outlook.

\section{Formalism}
\label{sec:Formalism}

\subsection{General setup in infinite volume}

We consider a system of two particles with reduced mass $\mu$ and relative
coordinate denoted $\vecr$ interacting via a finite-range, spherically
symmetric potential, \ie, for $R>0$ and
\begin{equation}
 \setball = \left\{ \vecx\in \RR^3: \abs{\vecx} \le R\right\} \,,
\end{equation}
it holds that
\begin{equation}
 V(\vecr,\vecr') = 0 \mathtext{if} \vecr,\vecr'\notin \setball \,,
\end{equation}
and $V$ depends only on the magnitude of $\vecr$ and $\vecr'$.
To be as general as possible, we allow $V$ to have a nonlocal form
and we emphasize that our main results to do not depend on the detailed
form of $V(\vecr,\vecr')$.
We furthermore note that although for convenience we assume a strict
finite range $R$ in the following, our results remain valid with
negligible corrections for short-range potentials that fall off faster
than any power law at large distances.

We write the Schrödinger equation for a state $\ket{\psi}$ as
\begin{equation}
 \hat{H}\ket{\psi} = E\ket{\psi} \,,
\end{equation}
with the Hamiltonian given by
\begin{equation}
 \hat{H}\psi(\vecr) = {-}\frac{1}{2\mu}\nabla^2 \psi(\vecr)
  + \int \ddr' \, V(\vecr,\vecr')\psi(\vecr') \,.
\end{equation}
We define $\ket{\psiinf}$ to be a solution with positive binding energy
$\Einfp$ such that
\begin{equation}
 \hat{H}\ket{\psiinf} = \Einf\ket{\psiinf} \,.
\end{equation}
For $\vecr \notin \setball$ the Hamiltonian simplifies to
\begin{equation}
 \hat{H}\psi(\vecr) \overset{\vecr \notin \setball}{=}
 {-}\frac{1}{2\mu}\nabla^2 \psi(\vecr)
\end{equation}
and we can write the asymptotic wave function as
\begin{equation}
 \psiinf(\vecr) \overset{\vecr \notin \setball}{\equiv} \psiinfa(\vecr)
 = {-}\ii^\ell \gamma h_\ell^{(1)}(i\kappa r)Y_\ell^m(\vecr/r) \,,
\label{qq:psi-infty-asymptotic}
\end{equation}
where $h^{(1)}_\ell$ is the spherical Hankel function of the first kind and
\begin{equation}
 \kappa^2=2\mu \Einfp \,.
\end{equation}
The asymptotic normalization coefficient (ANC) $\gamma$ is fixed via the
normalization condition $\braket{\psiinf}{\psiinf} = 1$
and Eq.~\eqref{eq:psi-infty-asymptotic}.
For later use we also define a state $\ket{\psiEasm}$ to be the purely asymptotic
form of a bound state with binding energy $E$, satisfying
\begin{equation}
 \hat{H}\psiEasm(\vecr) \overset{\vecr \notin \setball}{=} {-}E\psiEasm(\vecr) \,.
\end{equation}
We are not making any assumption here about the behavior of $\psiEasm(\vecr)$
for $\vecr\in\setball$ and just note that in all applications $\psiEasm$ will be
multiplied by an indicator function that is zero for $\vecr\in\setball$.

\subsection{Finite-volume wave function}

Now we consider the same system confined to a cubic periodic box of edge
length $L \gg R$.
The potential becomes periodic, taking the form
\begin{equation}
 V_L(\vecr, \vecr') = \sum_{\vecn} V(\vecr+\vecn L, \vecr'+ \vecn L) \,,
\end{equation}
and it satisfies
\begin{equation}
 V_L(\vecr, \vecr') = 0 \mathtext{if} \vecr\in\setasymptotic
 \mathtext{or} \vecr'\in\setasymptotic \,,
\end{equation}
where $\setasymptotic$, called the asymptotic domain, is defined as
\begin{equation}
 \setasymptotic = \left\{\vecx \in \RR^3: (\vecx + \vecn L) \notin B\
 \forall \vecn \in \ZZ^3 \right\} \,.
\end{equation}
The action of the finite-volume Hamiltonian on a generic state $\ket{\psi}$,
written in configuration space, is then
\begin{spliteq}
 \hat{H}_L\psi(\vecr) &= -\frac{1}{2\mu}\nabla^2 \psi(\vecr)
  + \int \ddr' \, V_L(\vecr,\vecr')\psi(\vecr')\\
 &= \hat{H}\psi(\vecr) + \sum_{\abs{\vecn}\ne 0}
  \int \ddr' \, V(\vecr + \vecn L,\vecr'+ \vecn L)\psi(\vecr') \\
 \overset{\vecr\in\setball\cup\setasymptotic}&{=} \hat{H}\psi(\vecr) \,.
\end{spliteq}

The exact finite-volume bound state $\psiL(\vecr)$ has an energy ${-}E(L)$
that depends on the size of the box.
We relate the infinite and finite volume-binding energies via
\begin{equation}
 E(L) = \Einfp + \Eshift \,,
\label{eq:Eshift-def}
\end{equation}
where $\Eshift$ is called the energy shift and has been investigated
extensively for various
systems~\cite{Luscher:1985dn,Konig:2011nz,Konig:2011ti,Konig:2017krd,Yu:2022nzm}.
In the derivation that follows we therefore treat $\Eshift$ as a known
quantity and we will frequently make use of the fact that
\begin{equation}
 \Eshift = \order{\ee^{{-}\kappa L}} \,.
\label{eq:Eshift-exp}
\end{equation}

The finite-volume wave function must be a solution to the finite-volume
Schrödinger equation with energy ${-}E(L)$ that obeys the periodic boundary
condition.
In the remainder of this section, we work out an ansatz for this periodic
finite-volume wave function that will form the basis for our derivation of
the radius volume dependence.

\paragraph*{Asymptotic solution}

Based on the observation that the infinite and finite-volume Hamiltonians are
equal in the asymptotic domain, we make the ansatz that an asymptotic
solution in finite volume is of the following form (analogous to what is
used in Refs.~\cite{More:2013rma,Furnstahl:2012qg,Furnstahl:2013vda}, based
on the ``linear energy method'' of Ref.~\cite{Djajaputra:2000aa}):
\begin{multline}
 \psiasm(\vecr) = \chi_\setasymptotic(\vecr)\Bigg[\psiinfa(\vecr) \\
 \null + \Eshift \dv{}{E} \psiEasm(\vecr)\eval_{E=\Einfp}\Bigg]
 + \order{\Eshift^2} \,.
\label{eq:psi-a}
\end{multline}
This means we consider the wave function as a function of the energy and relate
the volume dependence to an energy dependence via $E = E(L)$, allowing us to
Taylor-expand around infinite volume and keep the linear term explicitly.
We include $\chi_\setasymptotic(\vecr)$, the indicator function of
$\setasymptotic$, to conveniently set the state to zero outside of the
asymptotic domain.
If we act on $\ket{\psiasm}$ with $\hat{H}_L$ in the asymptotic domain, we find
\begin{widetext}
 \begin{spliteq}
 \hat{H}_L \psiasm(\vecr) \overset{\vecr\in\setasymptotic}{=}
 \hat{H} \psiasm(\vecr)&= \chi_\setasymptotic(\vecr)\left[\hat{H}\psiinfa(\vecr)
 + \Eshift \dv{}{E} \hat{H}\psiEasm(\vecr)\eval_{E=\Einfp}\right]
 + \order{\Eshift^2}\\
 &= \chi_\setasymptotic(\vecr) \left[\Einf \psiinfa(\vecr)
 - \Eshift \dv{}{E} E \psiEasm(\vecr)\eval_{E=\Einfp}\right]
 + \order{\Eshift^2}\\
 &=  \chi_\setasymptotic(\vecr) \left[\Einf \psiinfa(\vecr) - \Eshift
 \psiinfa(\vecr) - \Eshift \Einfp \dv{}{E} \psiEasm(\vecr)
 \eval_{E=\Einfp}\right]
 + \order{\Eshift^2}\\
 &=  -( \Einfp + \Eshift ) \chi_\setasymptotic(\vecr)\left[\psiinfa(\vecr)
 + \Eshift \dv{}{E} \psiEasm(\vecr)\eval_{E=\Einfp}\right] + \order{\Eshift^2} \\
 &=  \EL \psiasm(\vecr) + \order{\Eshift^2} \,.
\end{spliteq}
\end{widetext}
Therefore, to leading order, $\ket{\psiasm}$ solves the finite volume
Schrödinger equation with energy $\EL$ restricted to the asymptotic domain.

Now we intend to find a periodic solution based on $\ket{\psiasm}$.
We introduce a translation operator defined via
\begin{equation}
 \matrixelem*{\vecr}{\Trans(\vecn)}{\psi} = \psi(\vecr+\vecn L) \,,
\end{equation}
from which it follows that
\begin{equation}
 \matrixelem*{\psi}{\Trans^\dag(\vecn)}{\vecr} = \psi^*(\vecr + \vecn L) \,.
\end{equation}
It holds that
\begin{equation}
 \Trans^\dag(\vecn) = \Trans({-}\vecn)
\end{equation}
because
\begin{spliteq}
 \matrixelem*{\phi}{\Trans^\dag(\vecn)}{\psi}
 &= \int \ddr \, \phi^*(\vecr + \vecn L)\psi(\vecr) \\
 &= \int \ddr \, \phi^*(\vecr)\psi(\vecr - \vecn L) \\
 &= \matrixelem*{\phi}{\Trans(-\vecn)}{\psi} \,.
\end{spliteq}
Translation operators also have the property
\begin{equation}
 \Trans(\vecn)\Trans(\vecm) = \Trans(\vecn+\vecm) \,.
\end{equation}
Using translation operators we can construct the asymptotic finite-volume
wave function by adding shifted copies of $\ket{\psiasm}$ to satisfy the periodic
boundary condition.
This leads to:
\begin{equation}
 \ket{\psiLasm} = \sum_{\vecn} \Trans(\vecn)\ket{\psiasm}\,.
\end{equation}
Due to the linearity of the Schrödinger equation and the fact that the
finite-volume Hamiltonian commutes with our translation operators,
$\ket{\psiLasm}$ also satisfies the finite volume Schrödinger equation with
energy $\EL$, restricted to the asymptotic domain.

\paragraph*{Interior solution}

\begin{figure}
 \centering
 \includegraphics[width=\linewidth]{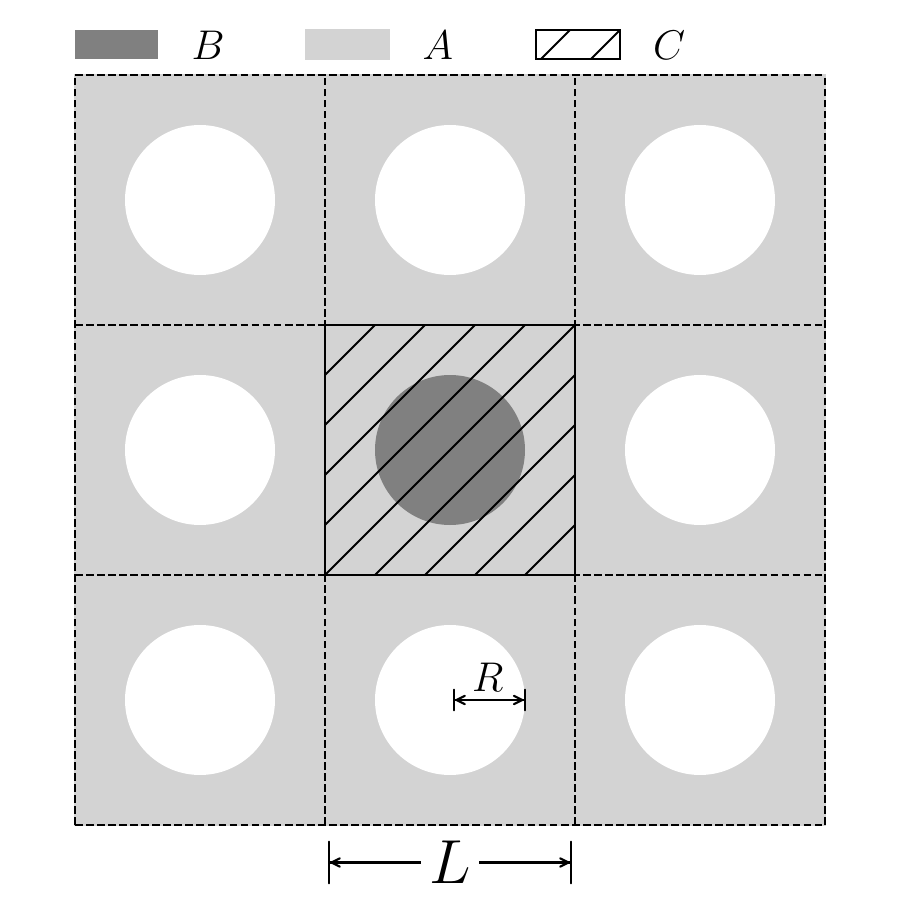}
 \caption{Illustration showing the sets $B$, $A$, and $C$ in a 2D analogy of
  the 3D scenario we consider.
  The central box $C$, cf.~Eq.~\eqref{eq:C}, is shown with neighboring
  periodic copies surrounding it.
  \label{fig:sets}
 }
\end{figure}

Now that we have found an asymptotic solution, we must find a solution for
$\vecr \notin \setasymptotic$.
We need only find an appropriate form for $\vecr \in \setball$, since periodic
copies of this set cover $A^C$, illustrated in Fig.~\ref{fig:sets}.

By our previous discussion, we know that $\psiinf(\vecr)$ is an approximate
solution for $\vecr\in\setball$ up to corrections of the order
$\order{\ee^{{-}\kappa L}}$.
Specifically, by Eqs.~\eqref{eq:Eshift-def} and~\eqref{eq:Eshift-exp} it holds
that
\begin{equation}
 \hat{H}_L \psiinf(\vecr) \overset{\vecr\in\setball}{=}
 \hat{H} \psiinf(\vecr) = {-}E(L)\psiinf(\vecr)
 + \order{\ee^{{-}\kappa L}} \,.
\end{equation}
Therefore we make the ansatz that an exact finite-volume solution for
$\vecr \in \setball$ is of the form
\begin{equation}
\label{eq:psiR-definition}
 \psiint(\vecr)
 = \chi_\setball(\vecr) \Big[\psiinf(\vecr) + \phi(\vecr)\Big] \,,
\end{equation}
where $\phi(\vecr)$ is some correction of order $\order{\ee^{{-}\kappa L}}$,
the detailed form of which we need not know.
This state must satisfy
\begin{equation}
 \hat{H}_L \psiint(\vecr) \overset{\vecr\in\setball}{=} \hat{H}\psiint(\vecr)
 = \EL\psiint(\vecr) \,.
\end{equation}
Expanding the left-hand side we get
\begin{equation}
 \hat{H}\psiint(\vecr) = \chi_\setball(\vecr)
 \Big[\Einf \psiinf(\vecr) + \hat{H}\phi(\vecr)\Big] \,.
 \end{equation}
Expanding also the right-hand side gives
\begin{equation}
 \EL\psiint(\vecr)
 = {-}\chi_\setball(\vecr)(\Einfp+\Eshift)
 \Big[\psiinf(\vecr) + \phi(\vecr)\Big] \,,
\end{equation}
and combining the two previous equations we obtain a differential equation
for $\phi(\vecr)$:
\begin{multline}
 \chi_\setball(\vecr)\hat{H}\phi(\vecr) = {-}\chi_\setball(\vecr)\Big[
 \Einfp \phi(\vecr) + \Eshift \psiinf(\vecr) \Big] \\
 + \order{\ee^{{-}2\kappa L}} \,.
\end{multline}
If we make the substitution $\ket{\phi} = \Eshift \ket{\varphi}$, we get
\begin{multline}
 \chi_\setball(\vecr)\hat{H}\varphi(\vecr)
 = {-}\chi_\setball(\vecr)\Big[\Einfp \varphi(\vecr) + \psiinf(\vecr)\Big] \\
 + \order{\ee^{{-}\kappa L}} \,.
\label{eq:varphi-def}
\end{multline}
Now we make the observation that the differential equation defining
$\varphi(\vecr)$ is independent of $L$ to leading order.
If the boundary conditions on $\varphi(\vecr)$ are also independent of $L$
to leading order then we can conclude that the $L$ dependence of $\phi(\vecr)$
must be limited to the factor of $\Eshift$ at this order.

We can fix two separate boundary conditions for $\varphi(\vecr)$.
For the first one, we use the fact that parity remains a good quantum number
in finite volume, so all states will have either even or odd parity.
For even parity states, the derivative of the wave function at the origin must
vanish along all three axes.
For odd parity states, the wave function must vanish at the origin.
Since $\psiint(\vecr)$ and $\psiinf(\vecr)$ have definite parities,
$\varphi(\vecr)$ must also have a definite parity by
Eq.~\eqref{eq:psiR-definition}, and therefore we obtain that either
$\varphi(\vecr)$ or its derivatives must vanish at the origin, and this
condition is independent of $L$.
For the second boundary condition we impose continuity between $\psiint$ and
$\psiLasm$:
\begin{widetext}
\begin{subalign}
 \psiint(\vecr)\eval_{\abs{\vecr}=R} &= \psiLasm(\vecr)\eval_{\abs{\vecr}=R}\\
 \psi_\infty(\vecr) + \Eshift\varphi(\vecr)\eval_{\abs{\vecr}=R}
 &= \sum_{\vecn} \Trans(\vecn)\psiasm(\vecr)\eval_{\abs{\vecr}=R}\\
 \psi_\infty(\vecr) + \Eshift \varphi(\vecr)\eval_{\abs{\vecr}=R}
 &= \sum_{\vecn}\Trans(\vecn)\bigg\{ \psiinfa(\vecr)
 +\Eshift \dv{}{E} \psiEasm(\vecr) \bigg\} \eval_{E=\Einfp,\abs{\vecr}=R} \\
 \psi_\infty(\vecr) + \Eshift \varphi(\vecr)\eval_{\abs{\vecr}=R}
 &= \psiinfa(\vecr)
 + \Eshift \dv{}{E} \psiEasm(\vecr)\eval_{E=\Einfp,\abs{\vecr}=R}
 + \sum_{\abs{\vecn}=1}\psiinfa(\vecn L) + \higherorder\\
 \varphi(\vecr)\eval_{\abs{\vecr}=R}
 &= \dv{}{E} \psiEasm(\vecr)\eval_{E=\Einfp,\abs{\vecr}=R}
 + \frac{1}{\Eshift}\sum_{\abs{\vecn}=1}\psiinfa(\vecn L)
 + \order{\ee^{(1-\sqrt{2})\kappa L}}\,.
\end{subalign}
\end{widetext}
For this boundary condition to be independent of $L$ to leading order,
it must be true that
\begin{equation}
\label{eq:varphi-bc}
 \frac{1}{\Eshift}\sum_{\abs{\vecn}=1}\psiinfa(\vecn L)
 = \text{const.} + \order{\ee^{(1-\sqrt{2})\kappa L}}.
\end{equation}
Although this is not true in general, it trivially holds for any odd parity
state since Eq.~\eqref{eq:varphi-bc} evaluates to zero.
We note that it also happens to hold for a number of even parity states, most
notably the S-wave.
In this paper we focus exclusively on S- and P-waves, for which this condition
is known to hold.
However, the following arguments work for any state as long as it can be shown
that Eq.~\eqref{eq:varphi-bc} holds.

Based on Eq.~\eqref{eq:varphi-def} and the boundary conditions, we conclude that
indeed $\varphi(\vecr)$ is independent of $L$ to leading order for at least
S- and P-wave states, and we write
\begin{equation}
 \psiint(\vecr) = \chi_\setball(\vecr)\Big[\psiinf(\vecr)
 + \Eshift \varphi(\vecr)\Big] + \higherorder \,,
\end{equation}
where all leading order $L$ dependence is now explicitly accounted for.
Just as before, we can make this solution periodic by adding shifted copies:
\begin{equation}
 \ket{\psiLint} = \sum_{\vecn} \Trans(\vecn)\ket{\psiint}\,.
\end{equation}

\paragraph*{Full construction}

Finally, we can join our two solutions to get the full finite-volume wave
function:
\begin{equation}
 \ket{\psiL} = \ket{\psiLasm} + \ket{\psiLint} = \sum_{\vecn}\Trans(\vecn)\left(
 \ket{\psiint} + \ket{\psiasm}\right) \,.
\end{equation}
$\ket{\psiLasm}$ and $\ket{\psiLint}$ are both periodic solutions to
\begin{equation}
 \hat{H}_L \ket{\psi} = {-}E(L)\ket{\psi}
\end{equation}
to order $\order{\ee^{{-}\sqrt{2}\kappa L}}$.
By linearity, $\ket{\psiL}$ must also be such a solution.
In addition, $\ket{\psiL}$ is periodic and continuous.

In the next section we will only need the form of $\psiL(\vecr)$ for $\vecr \in
\setcube$, where
\begin{equation}
 \setcube = \left({-}\frac{L}{2}, \frac{L}{2}\right)^3 \,.
\label{eq:C}
\end{equation}
Therefore, we drop the unused shifted copies of $\ket{\psiint}$ and rearrange:
\begin{equation}
 \ket{\psiL} \overset{\vecr\in\setcube}{=} \ket{\psiinf}
 + \Eshift\ket{\delta}
 + \sum_{\abs{\vecn}\ne 0} \Trans(\vecn)\ket{\psi_{a}}
 + \order{\ee^{{-}2\kappa L}} \,,
\end{equation}
where
\begin{equation}
 \ket{\delta} = \chi_\setball(\vecr)\ket{\varphi}
 + \chi_\setasymptotic(\vecr) \dv{}{E} \ket{\psiEasm}\eval_{E=\Einfp} \,.
\end{equation}

\subsection{Radius shift}

\subsubsection{Definition}

In general, we define the mean squared radius of a state as the expectation
value of the operator
\begin{equation}
 \hat{r}^2 = \frac{1}{N}\sum_{i=1}^N (\hat{\vecr}_i - \hat{\vecR})^2  \,,
\label{eq:r2-op}
\end{equation}
where $N$ is the number of particles, $\hat{\vecr}_i$ is the position
operator of the $i^\text{th}$ particle, and $\hat{\vecR}$ is the
center-of-mass position operator.
Since we work in coordinate representation we will drop the hats for
these operators in the following.
For two particles, we have
\begin{subalign}
 \vecr_1 &= \vecR + \frac{1}{2}\vecr \,, \\
 \vecr_2 &= \vecR - \frac{1}{2}\vecr \,,
\end{subalign}
where $\vecr = \vecr_2 - \vecr_1$ is the relative coordinate.
Plugging this into Eq.~\eqref{eq:r2-op}, we find that the mean
squared radius expectation value for two particles can be written in
terms of their relative coordinate simply as
\begin{equation}
 \expval{r^2} = \frac{1}{4}\expval{\vecr^2} \,.
\end{equation}\\
The finite-volume radius shift $\Delta\!\expval{r^2}(L)$ is defined as
\begin{equation}
 \Delta\!\expval{r^2}(L) = \expval{r^2}(L) - \expval{r^2_\infty} \,,
\end{equation}
where $\expval{r^2}(L)$ is
\begin{equation}
 \expval{r^2}(L) = \frac{1}{4}
 \frac{\matrixelem{\psi_L}{\vecr^2\chi_\setcube(\vecr)}{\psi_L}}
 {\matrixelem{\psi_L}{\chi_\setcube(\vecr)}{\psi_L}} \,.
\end{equation}
so $\Delta\!\expval{r^2}(L)$ can be written as
\begin{equation}
 \Delta\!\expval{r^2}(L) = \frac{1}{4} \frac{\matrixelem{\psi_L}{\vecr^2
 \chi_\setcube(\vecr)}{\psi_L}}
 {\matrixelem{\psi_L}{\chi_\setcube(\vecr)}{\psi_L}}
 - \expval{r^2_\infty} \,.
\label{eq:radius-shift-def}
\end{equation}
The matrix elements in the numerator and denominator of
Eq.~\eqref{eq:radius-shift-def} can all be written in the form
\begin{equation}
 \matrixelem{\psi_L}{\vecr^n \chi_\setcube(\vecr)}{\psi_L} \,,
\label{eq:general-radius-shift-integral}
\end{equation}
where $n=0$ in the denominator and $n=2$ in the numerator.

\subsubsection{Expansion}

Upon expanding the sums over shifted copies stemming from the definition of
$\ket{\psi_L}$ in Eq.~\eqref{eq:general-radius-shift-integral}, we get the
following terms
\begin{widetext}
\begin{multline}
 \matrixelem{\psi_L}{\vecr^n \chi_\setcube(\vecr)}{\psi_L}
 = \matrixelem*{\psiinf}{ \vecr^n \chi_\setcube(\vecr)}{\psiinf}
 + 2\Eshift
 \Rp\left[\matrixelem*{\psiinf}{\vecr^n \chi_\setcube(\vecr)}{\delta}\right] \\
 + \sum_{\abs{\vecn}\ne 0} 2\Rp\left[
  \matrixelem*{\psiinf}{\vecr^n \chi_\setcube(\vecr)\Trans(\vecn)}{\psiasm}
 \right]
 + \sum_{\substack{\abs{\vecn}\ne 0\\ \abs{\vecm}\ne 0}}
 \matrixelem*{\psiasm}{\Trans(-\vecn)\vecr^n
 \chi_\setcube(\vecr)\Trans(\vecm)}{\psiasm}
 + \order{\ee^{{-}2\kappa L}} \,.
\label{eq:psiL-expansion}
\end{multline}
We can add zero to the first term in the form of
\begin{equation}
 \matrixelem{\psiinf}{\vecr^n  \chi_\setcube(\vecr)}{\psiinf}
 = \matrixelem{\psiinf}{\vecr^n \chi_\setcube(\vecr)}{\psiinf}
 + \sum_{\abs{\vecn}\ne 0}
 \matrixelem{\psiinf}{\vecr^n \chi_\setcube(\vecr-\vecn L)}{\psiinf}
 -\sum_{\abs{\vecn}\ne 0}
 \matrixelem{\psiinf}{\vecr^n \chi_\setcube(\vecr-\vecn L)}{\psiinf} \,,
\end{equation}
\end{widetext}
which allows us to absorb the lone term into the first sum:
\begin{multline}
 \matrixelem{\psiinf}{\vecr^n \chi_\setcube(\vecr)}{\psiinf}
 = \sum_{\vecn}\matrixelem{\psiinf}{\vecr^n \chi_\setcube(\vecr - \vecn L)}
 {\psiinf} \\
 - \sum_{\abs{\vecn}\ne 0}
 \matrixelem{\psiinf}{\vecr^n\chi_\setcube(\vecr -\vecn L)}{\psiinf} \,.
\end{multline}
$\chi_\setcube(\vecr - \vecn L)$ simply sums to 1 over all $\vecn$, so
\begin{multline}
 \matrixelem{\psiinf}{\vecr^n \chi_\setcube(\vecr)}{\psiinf}
 = \matrixelem{\psiinf}{\vecr^n}{\psiinf}\\
 - \sum_{\abs{\vecn}\ne 0}
 \matrixelem{\psiinf}{\vecr^n\chi_\setcube(\vecr -\vecn L)}{\psiinf} \,.
\end{multline}
We write $\matrixelem{\psiinf}{\vecr^n}{\psiinf}$ as
$\expval{\vecr_\infty^n}$ and therefore we get:
\begin{multline}
 \matrixelem{\psiinf}{\vecr^n\chi_\setcube(\vecr)}{\psiinf}
 = \expval{\vecr_\infty^n} \\
 - \sum_{\abs{\vecn}\ne 0}
 \matrixelem{\psiinf}{\vecr^n\chi_\setcube(\vecr -\vecn L)}{\psiinf} \,.
\end{multline}

Expanding also the second term in Eq.~\eqref{eq:psiL-expansion}, we find
\begin{spliteq}
 2&\Eshift\,\Rp
 \Big[\matrixelem{\psiinf}{\vecr^n \chi_\setcube(\vecr)}{\delta}\Big] \\
 &= 2\Eshift\,\Rp\bigg[
  \matrixelem{\psiinf}{\vecr^n\chi_\setball(\vecr)}{\varphi} \\
 &\quad\null + \matrixelem{\psiinf}{\vecr^n \chi_{\setcube\cap\setasymptotic}(\vecr)
  \dv{}{E}}{\psiEasm}\!\eval_{E=\Einfp}
 \bigg] \,.
\label{eq:psiL-expansion-2nd}
\end{spliteq}
The first term in the square brackets on the right-hand side contains no $L$ dependence, and since we
do not explicitly know the form of $\psiinf(\vecr)$ for all $\vecr$, we choose to
parameterize the entire term by a constant $\beta_n'$.
Moreover, since $\psiEasm(\vecr)$ decays exponentially for large $r$, any $L$
dependence introduced by the second term will contribute only some decaying
exponential in $L$.
Since the right-hand side overall is already contains a factor of $\Eshift$,
the $L$ dependence from the second term turns out to be of higher exponential
order and can therefore be dropped.
Altogether, neither matrix element in Eq.~\eqref{eq:psiL-expansion-2nd} directly
contributes to the leading $L$ dependence and so we can combine them into
a single constant $\beta_n$:
\begin{multline}
 2\Eshift\,\Rp\matrixelem{\psiinf}{\vecr^n \chi_\setcube(\vecr)}{\delta}
 = \beta_n\Eshift \\
 + \higherorder \,.
\end{multline}

Turning to the third term in Eq.~\eqref{eq:psiL-expansion}, since
$\psiasm(\vecr)$ is always shifted by at least a distance $L$, and because the
second term in the definition of $\psiasm(\vecr)$, Eq.~\eqref{eq:psi-a}, is
already suppressed by a factor $\Eshift$, that part is overall beyond leading
exponential order and may be dropped.
Effectively, we may replace $\psiasm(\vecr)$ with $\psiinfa(\vecr)$, and we can
furthermore replace $\psiinf(\vecr)$ with the asymptotic form because it will
only ever be evaluated in the asymptotic region.
We therefore arrive at
\begin{multline}
 \sum_{\abs{\vecn}\ne 0}
 2\Rp\matrixelem*{\psiinf}{\vecr^n\chi_\setcube(\vecr) \Trans(\vecn)}{\psiasm} \\
 = \sum_{\abs{\vecn}=1}
 2\Rp\matrixelem*{\psiinfa}{\vecr^n \chi_\setcube(\vecr)\Trans(\vecn)
 \chi_\setasymptotic(\vecr)}{\psiinfa} \\
 + \higherorder \,,
\end{multline}
where we have also made use of the fact that only $\abs{\vecn} = 1$
terms contribute to leading exponential order.

Finally, for the fourth term in Eq.~\eqref{eq:psiL-expansion}, we can again note
that the second term of $\psiasm(\vecr)$ will contribute only beyond leading
exponential order since it is shifted by at least $L$ and suppressed
by $\Eshift$.
Therefore, we can replace $\psiasm(\vecr)$ with
$\chi_\setasymptotic(\vecr)\psiinfa(\vecr)$
and obtain
\begin{multline}
 \sum_{\substack{\abs{\vecn}\ne 0\\ \abs{\vecm}\ne 0}} \matrixelem*{\psiasm}
 {\Trans(-\vecn)\vecr^n \chi_\setcube(\vecr) \Trans(\vecm)}{\psiasm} \\
 = \sum_{\substack{\abs{\vecn}\ne 0\\ \abs{\vecm}\ne 0}}
 \matrixelem*{\psiinfa}{\Trans(-\vecn)\vecr^n
 \chi_{\setcube\cap\setasymptotic}(\vecr)\Trans(\vecm)}{\psiinfa} \\
 +\higherorder \,.
\end{multline}
Commuting the $\Trans(-\vecn)$ operator to the right furthermore gives
\begin{multline}
 \sum_{\substack{\abs{\vecn}\ne 0\\ \abs{\vecm}\ne 0}} \matrixelem*{\psiasm}
 {\Trans(-\vecn)\vecr^n\chi_\setcube(\vecr) \Trans(\vecm)}{\psiasm} \\
 = \sum_{\substack{\abs{\vecn}\ne 0\\ \abs{\vecm}\ne 0}}
 \bra{\psiinfa\strut}(\vecr - \vecn L)^n \chi_{\setcube\cap\setasymptotic}
 (\vecr - \vecn L) \\
 \null\times \Trans(\vecm - \vecn)\ket{\strut\psiinfa}
 +\higherorder \,.
\end{multline}
The only case in which this is of leading exponential order is when
$\vecn = \vecm$ and when their magnitude is 1, so
\begin{multline}
 \sum_{\substack{\abs{\vecn}\ne 0\\ \abs{\vecm}\ne 0}} \matrixelem*{\psiasm}
 {\Trans(-\vecn)\vecr^n\chi_\setcube(\vecr) \Trans(\vecm)}{\psiasm} \\
 = \sum_{\abs{\vecn} = 1} \matrixelem*{\psiinfa}{(\vecr - \vecn L)^n
 \chi_{\setcube\cap\setasymptotic}(\vecr - \vecn L)}{\psiinfa} \\
 + \higherorder \,.
\end{multline}
We can make the substitution
$\chi_{\setcube\cap\setasymptotic}(\vecr-\vecn L)
\to \chi_\setcube(\vecr-\vecn L)$ since including the shifted $\setball$ in the
integration domain only makes a less-than-leading order difference over the product
of shifted wave functions:
\begin{widetext}
\begin{equation}
 \sum_{\substack{\abs{\vecn}\ne 0\\ \abs{\vecm}\ne 0}} \matrixelem*{\psiasm}
 {\Trans(-\vecn)\vecr^n\chi_\setcube(\vecr) \Trans(\vecm)}{\psiasm}
 = \sum_{\abs{\vecn} = 1} \matrixelem*{\psiinfa}{(\vecr - \vecn L)^n
 \chi_\setcube(\vecr - \vecn L)}{\psiinfa}
 + \higherorder.
\end{equation}

Reassembling all the simplified terms back into Eq.~\eqref{eq:psiL-expansion},
we get:
\begin{multline}
\label{psiL simplified}
 \matrixelem{\psi_L}{\vecr^n \chi_\setcube(\vecr)}{\psi_L}
 = \expval{\vecr_\infty^n}
 + \beta_n\Eshift
 + \sum_{\abs{\vecn} = 1}\bigg \{
 \matrixelem*{\psiinfa}{((\vecr - \vecn L)^n - \vecr^n)
 \chi_\setcube(\vecr - \vecn L)}{\psiinfa} \\
 + 2\Rp\left[\matrixelem*{\psiinfa}{\vecr^n
 \chi_{\setcube\cap\setasymptotic}(\vecr)\Trans(\vecn)}{\psiinfa}\right]
 \bigg\}
 + \higherorder \,.
\end{multline}
Plugging this then into Eq.~\eqref{eq:radius-shift-def} and expanding to leading
exponential order, we arrive at
\begin{multline}
 \Delta\!\expval{r^2}(L) = \alpha \Eshift
 + \sum_{\abs{\vecn} = 1}\bigg[
 \matrixelem*{\psiinfa}{\frac{1}{4}(L^2 - 2\vecr\cdot\vecn L)
 \chi_\setcube(\vecr - \vecn L)}{\psiinfa} \\
 + \Rp\left\{\matrixelem*{\psiinfa}{\frac{1}{2}(\vecr^2 -4\expval{r^2_\infty})
 \chi_{\setcube\cap\setasymptotic}(\vecr)\Trans(\vecn)}{\psiinfa}\right\}
 \bigg]
 + \higherorder \,,
\label{eq:radius-shift}
\end{multline}
where $\alpha = \frac{1}{4}\beta_2 - \expval{r_\infty^2}\beta_0$ is a
parameter that must be fit by data.
\end{widetext}

\subsubsection{Simplification}

We now focus on simplifying both terms.
To that end, we write Eq.~\eqref{eq:radius-shift} in shorthand notation as
\begin{multline}
 \Delta\!\expval{r^2}(L) \\
 = \alpha \Eshift + \Rp \matrixelem{\psiinfa}{\hat{\eta}}{\psiinfa}
 + \higherorder \,,
\label{eq:Delta-r2-eta-hat}
\end{multline}
where
\begin{multline}
 \hat{\eta} = \sum_{\abs{\vecn} = 1}\bigg\{
 \frac{1}{4}(L^2-2\vecr\cdot\vecn L)\chi_\setcube(\vecr - \vecn L)\\
 + \frac{1}{2}\big[\vecr^2 -4\expval{r^2_\infty}\!\big]
 \chi_{\setcube\cap\setasymptotic}(\vecr)
 \Trans(\vecn) \bigg\} \,.
\label{eq:eta-hat}
\end{multline}
Note that we define $\hat{\eta}$ as an operator, but choose to write it
explicitly in terms of $\vecr$, with the understanding that this is equivalent
to writing $\hat{\vecr}$ when we specify that $\hat{\eta}$ is local in
coordinate space, $\matrixelem{\vecr'}{\hat{\eta}}{\vecr} =
\eta(\vecr)\delta^{(3)}(\vecr-\vecr')$.
Also note that the following manipulations of $\hat{\eta}$ are done with the
understanding that $\hat{\eta}$ will be evaluated between wave functions as in
Eq.~\eqref{eq:Delta-r2-eta-hat}.
All matrix elements are computed in configuration space, which is what we refer
to as ``integration'' in the following.

\begin{figure}
\centering
\includegraphics[width=\linewidth]{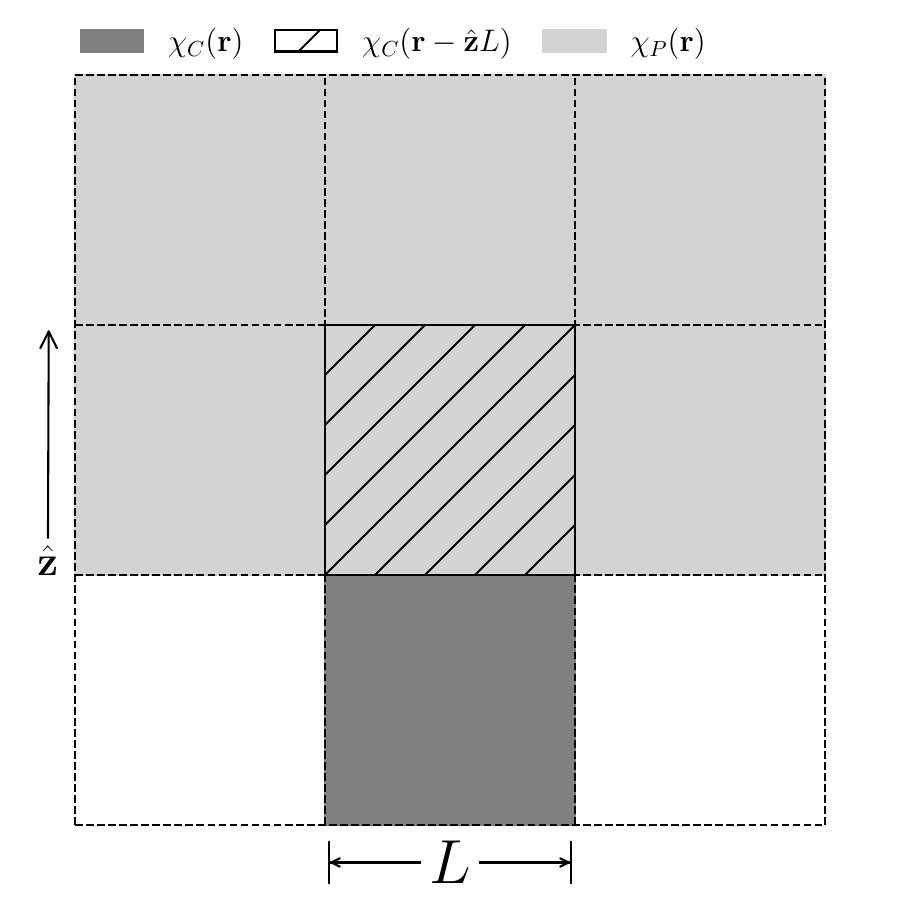}
\caption{Illustration showing the indicator functions of the sets $C$ and $P$ in a
 2D analogy.
 The central box is shown at the bottom with neighboring periodic copies
 surrounding it.
 Note that $P$ extends infinitely from the edges of the figure.
 \label{fig:p_set}
}
\end{figure}

We define the rotation operator $\Rot(\vecn)$ that maps the vector
$\vecn$ onto the direction of $\hatz$.
This definition does not uniquely describe a particular rotation, however, the
ambiguity does not matter for our purposes.
For example, we say that
\begin{equation}
 \Rot(\vecn)\Trans(\vecn)\Rot^\dag(\vecn) = \Trans(\hatz) \,.
\end{equation}
We note that
\begin{equation}
 \Rot^\dag(\vecn) = \Rot^{-1}(\vecn) \,.
\end{equation}
Because of this we can insert the identity $\Rot^\dag(\vecn)\Rot(\vecn)$
anywhere we would like.
Inserting this identity into $\hat{\eta}$ and commuting the operators to
opposite sides, we get
\begin{multline}
 \hat{\eta} = \sum_{\abs{\vecn} = 1}\Rot^\dag(\vecn)\bigg\{
 \frac{1}{4}(L^2 - 2z L) \chi_\setcube(\vecr - \hatz L) \\
 + \frac{1}{2}\big[\vecr^2 -4\expval{r^2_\infty}\!\big]
 \chi_{\setcube\cap\setasymptotic}(\vecr)
 \Trans(\hatz)\bigg\}\Rot(\vecn) \,.
\end{multline}
Since the real part will be taken at the end as per our previous manipulations,
we take the hermitian conjugate of the second term and then commute the translation
operator back to the right, leading to:
\begin{multline}
 \hat{\eta} = \sum_{\abs{\vecn} = 1}\Rot^\dag(\vecn)\bigg\{
 \frac{1}{4}(L^2 - 2z L)
 \chi_\setcube(\vecr - \hatz L) \\
 + \frac{1}{2}\big[(\vecr-\hatz L)^2 -4\expval{r^2_\infty}\!\big]
 \chi_{\setcube\cap\setasymptotic}(\vecr-\hatz L)
 \Trans(-\hatz)\bigg\}\Rot(\vecn).
\end{multline}
We can make the substitution $\chi_\setcube(\vecr - \hatz L) \to \chi_P(\vecr)$,
where
\begin{equation}
 P = \{\vecx \in \RR^3: \vecx\cdot \hatz > L/2\} \,,
\end{equation}
illustrated in Fig.~\ref{fig:p_set}.
This substitution does not cause a leading-order change, so we write
\begin{multline}
 \hat{\eta} = \sum_{\abs{\vecn} = 1}\Rot^\dag(\vecn)\bigg\{
 \frac{1}{4}(L^2 - 2z L)\chi_P(\vecr)\\
 + \frac{1}{2}\big[(\vecr-\hatz L)^2 -4\expval{r^2_\infty}\!\big]
 \chi_{P\cap\setasymptotic}(\vecr)\Trans(-\hatz)\bigg\}\Rot(\vecn) \,.
\end{multline}

We may furthermore insert the function $\chi_\setasymptotic(\vecr)$ into the
first term without introducing a leading-order change.
This makes the indicator function conveniently identical for both terms:
\begin{multline}
 \hat{\eta} = \sum_{\abs{\vecn} = 1}
 \Rot^\dag(\vecn)\chi_{P\cap\setasymptotic}(\vecr)\bigg\{
 \frac{1}{4}(L^2 - 2z L)\\
 + \frac{1}{2}\big[(\vecr-\hatz L)^2 -4\expval{r^2_\infty}\!\big]
 \Trans(-\hatz)\bigg\}\Rot(\vecn) \,.
\end{multline}
We can write this now as
\begin{equation}
 \hat{\eta} = \sum_{\abs{\vecn} = 1}\Rot^\dag(\vecn)\hat{\xi}\Rot(\vecn) \,,
\end{equation}
where
\begin{multline}
 \hat{\xi} = \chi_{P\cap\setasymptotic}(\vecr)\bigg\{\frac{1}{4}(L^2-2z L)\\
 + \frac{1}{2}((\vecr-\hatz L)^2 -4\expval{r^2_\infty})\Trans(-\hatz)\bigg\} \,.
\end{multline}
Finally, we arrive at the final simplified form:
\begin{multline}
\label{eq:analytic-radius-shift}
 \Delta\!\expval{r^2}(L) = \alpha \Eshift \\
 +\Rp\left( \sum_{\abs{\vecn}=1} \matrixelem{\Rot(\vecn)\psiinfa}
 {\hat{\xi}}{\Rot(\vecn)\psiinfa} \right)\\
 + \higherorder \,.
\end{multline}
Depending on the actual (cubic) symmetry properties of $\psiinfa$, some of the
terms in the sum may not need to be computed.
For example, for the finite-volume analog of an S-wave state (discussed in more
detail below), $\Rot(\vecn)\psiinfa(\vecr) = \psiinfa(\vecr)$ for all
rotations $\Rot(\vecn)$, so all the terms
in the sum are identical and only one needs to be evaluated explicitly.

\subsubsection{Coordinate transformation}

Now we present a coordinate system in which the quantity
$\bra{\psi}\hat{\xi}\ket{\psi}$ is relatively straightforward to compute in
closed form.
We use a two-center bispherical coordinate system parameterized by $r$, $u$, and
$\phi$, where $r$ is the distance from the origin, $u$ is the distance from the
origin of the neighboring periodic box in the $\hatz$ direction, and $\phi$ is
the azimuthal angle about the $z$ axis.
The valid set of coordinates is given by
\begin{equation}
 \{(r,u,\phi)\in [0,\infty)\times [0,\infty)\times [0,2\pi): r+u \ge L\} \,.
\end{equation}
For reference, we note that the transformation to Cartesian coordinates is
provided by
\begin{align}
 x &=  \sqrt{r^2-\frac{\left(L^2+r^2-u^2\right)^2}{4 L^2}}\cos \phi \,, \\
 y &=  \sqrt{r^2-\frac{\left(L^2+r^2-u^2\right)^2}{4 L^2}}\sin \phi \,, \\
 z &= \frac{L^2+r^2-u^2}{2 L} \,.
 \end{align}
 The volume element in the $(r,u,\phi)$ coordinate system has the simple form
 \begin{equation}
 \ddr = \frac{ru}{L}\dd{r}\dd{u}\dd{\phi} \,.
 \end{equation}
The main advantage of using this coordinate system is that
\begin{equation}
 \Trans(-\hatz)r = \abs{\vecr - \hatz L} = u \,,
\end{equation}
so integrals over shifted radial functions are just as simple as integrals over
unshifted radial functions.
The trade-off, however, is that the bounds of the integration domain become more
complicated, especially given that not all coordinate values are valid.
With this in mind, the bounds of integration with the indicator function
$\chi_{P\cap\setasymptotic}(\vecr)$ become
\begin{multline}
\label{eq:matrix-element-integrals}
 \int \chi_{P\cap\setasymptotic}(\vecr)\dots \ddr
 = \int_{R}^{L/2}\int_{L-u}^{L+u}\int_0^{2\pi} \dots
 \frac{ru}{L}\dd{\phi}\dd{r}\dd{u} \\
 + \int_{L/2}^{\infty}\int_{u}^{\infty}\int_0^{2\pi}
 \dots \frac{ru}{L}\dd{\phi}\dd{r}\dd{u}.
\end{multline}
In the $(r,u,\phi)$ coordinate system, $\hat{\xi}$ has the form
\begin{multline}
 \hat{\xi} = \chi_{P\cap\setasymptotic}(\vecr)\bigg\{\frac{1}{4}(u^2-r^2)
 + \frac{1}{2}(u^2-4\expval{r^2_\infty})\Trans(-\hatz)\bigg\} \,.
\label{eq:xi-final}
\end{multline}

The coordinate system described above allows us to obtain closed-form
expressions for the finite-volume radius shift for certain important cases,
which we present in the following section.
We use this formalism to derive explicit analytical expressions for the radius shift for S- and P-wave states shown in Sec.~\ref{sec:Explicit}, with details of the calculation presented in the appendix.

\section{Results}
\label{sec:Results}

\subsection{Broken spherical symmetry}

As already alluded to before, spherical symmetry is broken by confining the
system to a periodic cubic box, and as a consequence angular momentum $\ell$
is not a good quantum number anymore.
The relevant spatial symmetry is instead described by the group of rotations
that leave a cube invariant.
The structure of this group is well known and has been discussed, for example,
in Ref.~\cite{Johnson:1982yq}.
Angular-momentum multiplets in general break up into irreducible representations
of the cubic group, of which there are overall five different ones,
denoted as $\Gamma = A_1, A_2, E, T_1, T_2$, with dimensions $1,1,2,3,3$, respectively.
It is typically a good assumption
to identify S-wave ($\ell=0$) states in infinite volume with $A_1^+$ cubic
states, and P-wave states with the $T_1^-$ representation, where the superscript
indicates positive or negative parity.\footnote{
Higher angular momenta $\ell$ contribute to both cubic multiplets, but there are significant
gaps.
As discussed in Ref.~\cite{Johnson:1982yq}, $A_1^+$ receives
contributions from $\ell = 0,4,6,8,\ldots$,
while for $T_1^-$ the sequence is
$\ell=1,3,4,5,\ldots$.}

\subsection{Explicit formulae}
\label{sec:Explicit}

We present here analytic forms of the radius shift for $\ell=0$ and $\ell=1$
contributions to the $A_1^+$ and $T_1^-$ cubic representations, respectively.
After calculating the radius shift for S-wave and P-wave states using the
method presented above, we find that
\begin{multline}
 \Delta\!\expval{r^2}_0^{A_1^+}(L) = \\
 \abs{\gamma}^2 \ee^{{-}\kappa L}
 \left(\frac{L^2}{4\kappa}+\frac{3 \left(1-8 \kappa^2
 \expval{r^2_\infty}\right)}{8\kappa^3} + \frac{a}{\kappa^4  L}\right) \\
 + \frac{3}{16} \abs{\gamma}^2 L^3 \text{Ei}({-}\kappa L )
 + \higherorder
\label{eq:radius-shift-A1p}
\end{multline}
and
\begin{multline}
 \Delta\!\expval{r^2}_1^{T_1^-}(L) = \\
 \abs{\gamma}^2 \ee^{{-}\kappa  L}
 \left(-\frac{L^2}{4 \kappa }+\frac{3 \left(5+8 \kappa ^2
 \expval{r^2_\infty}\right)}{8\kappa^3}+\frac{a}{\kappa^4 L}\right) \\
 +\frac{3}{16\kappa^2}\abs{\gamma}^2 L(8-\kappa^2 L^2)\text{Ei}({-}\kappa L)
 + \higherorder \,,
\label{eq:radius-shift-T1m}
\end{multline}
where $a$ is a dimensionless fit parameter.
The details of how we arrived at these expressions are given in the
appendix.
We make the following observations:
\begin{enumerate}[wide,labelwidth=!,labelindent=0pt]
\item S-wave representations are one dimensional, and the same is true for
$A_1^+$.
All three basis states for $\ell=1$ (and therefore for $\Gamma=T_1^-$ in
the P-wave approximation) have the same radius shift since they are all just
different rotations of essentially the same degenerate state.
\item In general, the radius-shift formula will have at least two
additional fit parameters compared to the energy shift.
The first of these is the $\alpha$ introduced in Eq.~\eqref{eq:radius-shift},
while the second is $R$, the upper bound of the interior integration domain
that enters via the indicator function $\chi_{P\cap\setasymptotic}$ in Eq.~\eqref{eq:xi-final},
which is in general unknown if the interaction does not have a strict finite
range.
Remarkably, however, the S- and P-wave radius shift formulae still feature only
\emph{one} additional fit parameter compared to the energy shift since we were
able to absorb all $R$ and $\alpha$ dependence into a constant $a$, as
shown in the appendix.
\item We also note that the S- and P-wave radius shifts are exactly negatives
of each other up to order
$\order{\ee^{{-}\kappa L}\times L^0} = \order{\ee^{{-}\kappa L}}$.
That is, the S- and P-wave radius shifts are
\begin{align*}
 \Delta\!\expval{r^2}_0^{A_1^+}(L) &= \frac{\abs{\gamma}^2 \ee^{{-}\kappa L}}
 {16 \kappa ^2}\qty(\kappa L^2 + 3L + \order{L^0}) \\
 \Delta\!\expval{r^2}_1^{T_1^-}(L) &= -\frac{\abs{\gamma}^2 \ee^{{-}\kappa L}}
 {16 \kappa ^2}\qty(\kappa L^2 + 3L + \order{L^0}) \,.
\end{align*}
This finding is similar to what has been found for the finite-volume energy
shifts~\cite{Konig:2011ti}.
\item From the correlation between binding energies and mean-squared radii that
in infinite-volume usually means that more deeply bound states become more
compact, combined with the known volume dependence of the binding
energy,~\cite{Konig:2011nz,Konig:2011ti} one might intuitively expect the
leading radius corrections to have exactly the opposite signs of what we
found here.
S-wave bound states become more deeply bound in finite volume, so the naive
intuition would be that their radii would \emph{decrease}, and vice versa for
P-wave states.
However, the behavior we derived here can in fact be related to
how the finite volume affects the wave functions, similar to the intuitive
argument that explains the sign of the energy
shift~\cite{Konig:2011nz,Konig:2011ti}:
Since $A_1^+$ S-wave states have even parity, the derivative of the wave
function across the boundary of the box must be zero.
This means that the finite-volume wave function can have a larger-magnitude
tail near the boundary of the box than the corresponding infinite-volume
wave function at the same distance.
Since the mean-squared radius defined as the expectation value of $r^2$
relatively emphasizes contributions from large distances, overall the radius
can increase in finite volume.
For $T_1^-$ P-wave states on the other hand, odd parity forces the wave
function to zero at the boundary, compressing the wave function profile
and therefore leading to a smaller radius compared to infinite volume.
\end{enumerate}

\subsection{Numerical checks}

Part of this work is determining the optimal strategy for using the shift
formulae for practical radius extrapolations.
For this purpose, we assume that we are dealing with a finite-volume simulation
that provides both energies and wave functions for the states of interest, such
as a straightforward lattice discretization of the Hamiltionian or a discrete
variable representation (DVR) based on plane-wave states, an efficient few-body
implementation of which has been discussed in
Refs.~\cite{Klos:2018sen,Konig:2020lzo,Konig:2022cya}.
Since we have access to the energy data, it makes sense to use that first to
extract the binding momentum $\kappa$ and the ANC $\gamma$.
Once $\kappa$ and $\gamma$ have been determined, they can be used as fixed
parameters in the radius volume dependence, leaving only two parameters still to
be fit, $\expval{r^2_\infty}$ and $a$.

Determining a particular ``best'' fitting algorithm is difficult due to the
unknown higher-order terms and the exponential form of the volume dependence.
One option, employed in much of the FV bound-state literature cited previously,
is to fit the data on a logarithmic scale and focus on the large-volume
region where the higher-order corrections are smaller.
However, fitting on a logarithmic scale introduces several complications.
Firstly, in order to obtain a simple form, one generally needs to subtract the
infinite-volume value from the data.
While this can be done relatively easily for the binding energy in some
cases,\footnote{%
For $A_1^+$ and $T_1^-$ states without Coulomb interaction, the leading volume
dependence is a pure exponential, so one can determine the infinite-volume
energy by demanding that the volume dependence is linear on an (approximately
scaled) logarithmic scale, as done for example in Ref.~\cite{Konig:2017krd}.`}
in general the infinite-volume value is one of the fit parameters,
so we do not know its value before performing the fit.
Even after getting past that problem, logarithmic scales can make it very
difficult to determine constant terms, such as $\expval{r^2_\infty}$ in
\begin{equation}
 \expval{r^2}(L) = \expval{r^2_\infty} + \Delta \expval{r^2}(L) \,,
\label{eq:radius-shift-fit}
\end{equation}
cf.~Eq.~\eqref{eq:radius-shift}, because the logarithm of the right-hand side
diverges near the correct value, \ie, when the fit is near optimal.
Since the residuals may not reflect the true quality of the
fit, this fitting method tends to be unstable.

The method we propose assumes that the only source of uncertainty comes from
the unknown higher-order terms.
Therefore, it makes sense to simply minimize the residuals on a linear scale
weighted by the inverse absolute value of those higher order terms.
Since of course we do not know the exact form of the higher-order terms, we
merely assume that they scale appropriately as $\ee^{{-}\sqrt{2}\kappa L}$.
To illustrate this method with concrete examples, we perform fits using a
weighted least-squares algorithm where the weights are assigned as described
above.
Specifically, we apply this procedure to perform fits of the form
\begin{equation}
 E(L) = \Delta E(L) + E_\infty
\end{equation}
for the energy, and as in Eq.~\eqref{eq:radius-shift-fit} for the radius.

For our numerical simulation we use the DVR framework of
Ref.~\cite{Klos:2018sen} and as concrete interaction we use attractive
local Gaussian potentials of the form
\begin{equation}
 V(r) = V_0 \exp({-}\left(\frac{r}{R_0}\right)^2) \,,
\end{equation}
with parameters $V_0<0$ and $R_0>0$.
This interaction does not have a strict finite range $R$ as assumed for
convenience in the derivation of the radius volume dependence, but corrections
stemming from the Gaussian tails of the potential can generally be neglected.

\begin{figure}[tbhp]
\centering
\includegraphics[width=\linewidth]{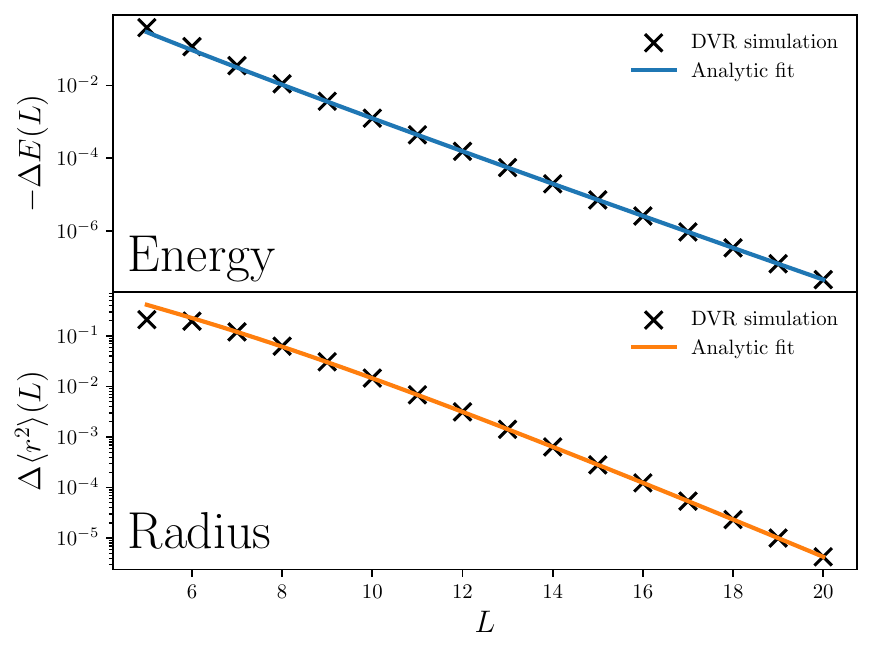}
\caption{Volume dependence for the energy and mean squared radius of an S-wave
 state using a Gaussian potential ($R=2$, $V_0={-}3$).
 Quantities are reported in natural units with the particle mass set to 1.
 The energy and radius shifts from a numerical simulation and the analytic
 fit are plotted in the upper and lower panels, respectively.
 \label{fig:s-wave}
}
\end{figure}

\begin{figure}[tbhp]
\centering
\includegraphics[width=\linewidth]{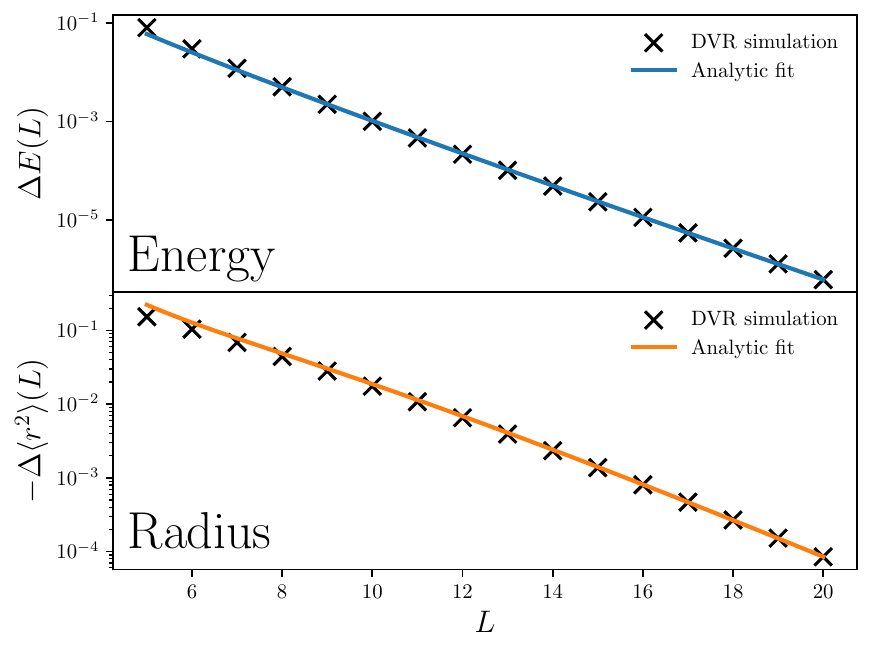}
\caption{Volume dependence for the energy and mean squared radius of a P-wave
 state using a Gaussian potential ($R=1$, $V_0={-}14$).
 Quantities are reported in natural units with the particle mass set to 1.
 The energy and radius shifts from a numerical simulation and the analytic
 fit are plotted  in the upper and lower panels, respectively.
 \label{fig:p-wave}
}
\end{figure}

Example fits for S- and P-wave states are shown in Fig.~\ref{fig:s-wave} and
Fig.~\ref{fig:p-wave}, respectively.
Even though the fits were performed on a linear scale, as described above, they
look excellent on the logarithmic scale that we chose to improve the display
and highlight the excellent agreement of the numerical simulation with our
predictions.
This is because of the exponential weighting.

The quality of the fit for the radius is particularly good given that $\kappa$
and $\gamma$ were predetermined by the energy fit and the fit parameter $a$
has very little influence on the shape of the curve.
We see the analytic fit significantly deviating from the simulation data
only in very small volumes due to the higher order terms and due to
violating of the condition $L\gg R$.
The latter is most likely the dominant reason because we observe that the
deviation occurs over approximately the same volume range for both the energy
and the radius, with comparable magnitude.


A quantitative overview of the radius extrapolations we can get using this
method is shown in Tables~\ref{tab:S fit ranges} and~\ref{tab:P fit ranges}.
For comparison, continuum results were calculated as reference by numerically
solving the radial equation for the system via the shooting method and
evaluating the mean squared radius using the wave functions obtained in that
manner.
The radius extrapolations perform well over a variety of volume ranges and
the extrapolated radius is consistently more accurate (compared to the reference results) than the radius from the
largest simulated volume.
We note that the uncertainty in the extrapolated radius extracted from
the fits is only a lower bound for the true theoretical uncertainties.
A more sophisticated approach would propagate the uncertainties in
$\kappa$ and $\gamma$ from the energy fits and include include also the
systematic uncertainty stemming from omitted higher order terms in the
radius volume dependence.

\begin{table}
\centering
\renewcommand{\arraystretch}{1.1}
\begin{tabular}{ccl}
 \toprule
 \multicolumn{3}{c}{Fit Range: $L=6\cdots L_\mathrm{max}$} \\
 \hline
 $L_\mathrm{max}$
 & $\expval{r^2}(L_\mathrm{max})$
 & \multicolumn{1}{c}{$\langle r^2_\infty\rangle_\mathrm{fit}$}
 \rule{0pt}{1.15em}\\[0.15em]
 \hline
 $~~10~~$ & $0.77457786831802$ & $0.7625(3)$ \\
 $12$ & $0.76302346893672$ & $0.76022(3)$ \\
 $14$ & $0.76051685324793$ & $0.759917(3)$ \\
 $16$ & $0.76000414620613$ & $0.7598839(3)$ \\
 $18$ & $0.75990366351178$ & $0.75988064(2)$ \\
 $20$ & $0.75988461024428$ & $0.759880344(2)$ \\
 \hline
 \multicolumn{3}{c}{Continuum: $0.75988031$} \\
 \bottomrule
\end{tabular}
\caption{Fit results for $\langle r^2_\infty\rangle$ over different volume
 ranges for an S-wave state ($R=2,V_0={-}3$) compared to the mean squared
 radius at the largest volume in the fit region.
 Quantities are reported in natural units with the particle mass set to 1.
 \label{tab:S fit ranges}
}
\end{table}
\begin{table}
\centering
\renewcommand{\arraystretch}{1.1}
\begin{tabular}{ccl}
 \toprule
 \multicolumn{3}{c}{Fit Range: $L=6\cdots L_\mathrm{max}$}\\
 \hline
 $L_\mathrm{max}$
 & $\expval{r^2}(L_\mathrm{max})$
 & \multicolumn{1}{c}{$\langle r^2_\infty\rangle_\mathrm{fit}$}
 \rule{0pt}{1.15em}\\[0.15em]
 \hline $10$ & $0.578866122009667$ & $0.5927(3)$\\
 $~~12~~$ & $0.589849429592643$ & $0.59527(8)$\\
 $14$ & $0.594046766392319$ & $0.59607(2)$\\
 $16$ & $0.595575413626522$ & $0.596305(5)$\\
 $18$ & $0.596112250896048$ & $0.596366(1)$\\
 $20$ & $0.596295400058964$ & $0.5963814(2)$\\
 \hline
 \multicolumn{3}{c}{Continuum: $0.5963857$}\\
 \bottomrule
\end{tabular}
\caption{Fit results for $\langle r^2_\infty\rangle$ over different volume
 ranges for a P-wave state ($R=1,V_0={-}14$) compared to the mean squared
 radius at the largest volume in the fit region.
 Quantities are reported in natural units with the particle mass set to 1.
 \label{tab:P fit ranges}
}
\end{table}

\section{Summary and outlook}
\label{sec:Conclusion}

We have studied the leading volume dependence for the mean squared radius
of bound states of two point particles in a finite periodic box.
Using an ansatz for the wave function in finite volume and a sequence of
systematic simplifications, we derived a general formula for the finite-volume
correction to the radius expectation value.
With the help of a carefully constructed coordinate system, we were able to
evaluate this general expression and obtain closed-form expressions for the
important cases of S- and P-wave states falling within the $A_1^+$ and $T_1^-$
irreducible representations of the cubic group in finite volume.
These expressions are to a large extent informed by the volume dependence of the
energy and involve a surprisingly small number of additional parameters that
need to be fit to numerical simulation data.
As part of this work we have performed such numerical simulation using Gaussian
model potentials and found excellent agreement of our analytic results with
calculations.

Our results constitute important progress towards obtaining precise predictions
from finite-volume simulations for observables beyond binding energies for
quantum systems such as atomic nuclei.
While we have studied here explicitly the mean squared radius, our method of
constructing an ansatz for the periodic finite-volume wave function without
explicit knowledge of the short-distance behavior, and subsequently evaluating
matrix elements based on this ansatz, provides a recipe for deriving the volume
dependence of other static properties, such as for example quadrupole moments.

An important next step towards implementing radius extrapolations in practical
applications will be the extension of our findings to bound states comprised of
more than two particles.
Guidance for such work can be provided by the formalism that derived the
binding-energy volume dependence for arbitrary cluster
states~\cite{Konig:2017krd}.
Moreover, recent work that studies charged-particle bound states in periodic
boxes~\cite{Yu:2022nzm} can inform the extension of our method to such systems.

Finally, it is worth noting that radii of atomic nuclei are typically measured
using electromagnetic scattering processes.
Specifically, charge radii can be inferred from the slope of the so-called
charge form factor $F_C(\vecq^2)$, where $\vec{q}$ is the momentum transferred
to the nucleus by virtual-photon exchange, in the limit $\vecq^2\to0$.
Matter radii can then be further estimated from the measured charge radii.
For theory, it is desirable to follow an analogous procedure, which compared to
evaluating the expectation value of $r^2$ can ensure consistency with the
experimental determination and in particular take into account a systematic
expansion of the electromagnetic current operator.
Following this approach in finite volume requires understanding the volume
dependence of $F_C(\vecq^2)$, which can be informed by the results presented
in this paper.

\begin{acknowledgments}
We thank Lex Kemper and Dean Lee for useful discussions.
We furthermore acknowledge work by Brandon Holton on a precursor of
the research presented in this paper.
This work was supported by the National Science Foundation under Grant
No. PHY--2044632 as well as by the U.S.\ Department of Energy (DE-SC0024520
-- STREAMLINE Collaboration and DE-SC0024622).
This material is based upon work supported by the U.S. Department of Energy,
Office of Science, Office of Nuclear Physics, under the FRIB Theory Alliance,
award DE-SC0013617.
\end{acknowledgments}

\medskip
\appendix

\section{Detailed derivation of the final radius-shift expressions}

We calculate here the explicit final expressions for the finite-volume radius
shifts given in the main text.

\subsection{S-wave evaluation}
\label{sec:S-wave-calc}

For an S-wave state, the infinite-volume wave function has the form
\begin{equation}
 \psiinfa(r,\theta,\phi)
 = \frac{\gamma}{\sqrt{4\pi}} \frac{\ee^{-\kappa r}}{r}
\end{equation}
for large $r$.
Since the S-wave state is rotationally symmetric and can be chosen to be real,
we can simplify Eq.~\eqref{eq:analytic-radius-shift} to
\begin{multline}
 \Delta\!\expval{r^2}_0^{A_1^+}\!\!\!(L) \\
 = \alpha \Eshift
 + 6\,\matrixelem*{\psiinfa}{\hat{\xi}}{\psiinfa}
 + \higherorder \,.
\end{multline}
We can also insert the known form of the S-wave energy shift to get
\begin{multline}
 \Delta\!\expval{r^2}_0^{A_1^+}\!\!\!(L) \\
 = {-}\frac{3\abs{\gamma}^2\alpha}{\mu L} \ee^{-\kappa L}
 + 6\,\matrixelem*{\psiinfa}{\hat{\xi}}{\psiinfa}
 \null + \higherorder \,.
\end{multline}
Writing out the integrand that appears in the  evaluation of the matrix element,
we get
\begin{multline}
 \psiinfa^*(\vecr)\hat{\xi}\psiinfa(\vecr) \\
 = \chi_{P\cap\setasymptotic}(\vecr)
 \frac{\abs{\gamma}^2 \ee^{{-}\kappa (r+u)}}{16\pi r^2 u}
 \bigg(2 r \left(u^2-4\expval{r^2_\infty}\right)\\
 + u (u-r) (r+u) \ee^{\kappa (u-r)}\bigg) \,,
\end{multline}
noting that both $u$ and $r$ will be integrated over as described in the
main text.
We now perform the integrals in Eq.~\eqref{eq:matrix-element-integrals} with
this  integrand and drop higher-order terms:
\begin{widetext}
\begin{multline}
 \matrixelem*{\psiinfa}{\hat{\xi}}{\psiinfa}
 = \frac{\abs{\gamma}^2}{32} L^3 \Ei({-}\kappa L)
 + \abs{\gamma}^2\ee^{-\kappa L} \bigg(
  \frac{L^2}{24 \kappa }
  + \frac{\left(1-8 \kappa ^2 \expval{r^2_\infty}\right)}{16 \kappa ^3} \\
  - \frac{\ee^{-2\kappa R} \left(
   \ee^{2\kappa R} \left(
    4\kappa^2 \left(\kappa  R^3+2\expval{r^2_\infty} (3-6\kappa R)\right) - 3
   \right)
   + 6\kappa \left(\kappa  R^2+R-4\kappa\expval{r^2_\infty}\right)
   + 3
  \right)}{48\kappa^4 L}
 \bigg) + \higherorder \,.
\end{multline}
Putting everything back together we get
\begin{multline}
 \Delta\!\expval{r^2}_0^{A_1^+}\!\!\!(L)
 = \frac{3}{16}\abs{\gamma}^2 L^3 \Ei(-\kappa L)
 + \abs{\gamma}^2\ee^{{-}\kappa L}\bigg(
  \frac{L^2}{4 \kappa }
  + \frac{3\left(1-8 \kappa ^2 \expval{r^2_\infty}\right)}{8 \kappa ^3}
  - \frac{3\alpha}{\mu L} \\
  \null - \frac{\ee^{-2\kappa R} \left(
   \ee^{2\kappa R}\left(
    4\kappa^2
    \left(\kappa R^3+2\expval{r^2_\infty} (3-6 \kappa  R)\right)-3
   \right) +6 \kappa \left(
    \kappa R^2+R - 4\kappa\expval{r^2_\infty}
   \right)
   + 3
  \right)}{8\kappa^4 L}
 \bigg) + \higherorder \,.
\end{multline}
\end{widetext}
We can absorb many of the constants that appear in this expression into a
single constant $a$.
Doing that, we arrive at Eq.~\eqref{eq:radius-shift-A1p} in the main text.

\subsection{P-wave evaluation}
\label{sec:P-wave-calc}

As stated in the main text, all three basis states for $\ell=1$ and
$\Gamma=T_1^-$ have the same radius shift since they are all just different
rotations of essentially the same degenerate state.
It therefore suffices to consider just one of the wave functions in the multiplet, the asymptotic form of which we can write as
\begin{equation}
 \psiinfa(r,\theta,\phi) = \sqrt{\frac{3}{4\pi }}
 \frac{\gamma \ee^{{-}\kappa r} (\kappa r+1) \cos\theta }{\kappa  r^2}.
\label{eq:psi-P-asympy}
\end{equation}
We begin again with Eq.~\eqref{eq:analytic-radius-shift}.
Expanding the sum and plugging in the known form of the P-wave energy shift,
we get
\begin{widetext}
\begin{multline}
 \Delta\!\expval{r^2}_1^{T_1^-}\!\!\!(L) = \frac{3\abs{\gamma}^2\alpha}{\mu L}
 \ee^{{-}\kappa L} +\Rp\bigg[
 \matrixelem*{\Rot(\hatx)\psiinfa}{\hat{\xi}}{\Rot(\hatx)\psiinfa}
 + \matrixelem*{\Rot(\haty)\psiinfa}{\hat{\xi}}{\Rot(\haty)\psiinfa} \\
 \null + \matrixelem*{\Rot(\hatz)\psiinfa}{\hat{\xi}}{\Rot(\hatz)\psiinfa}
 + \matrixelem*{\Rot(-\haty)\psiinfa}{\hat{\xi}}{\Rot(-\haty)\psiinfa}
 \null + \matrixelem*{\Rot(-\hatx)\psiinfa}{\hat{\xi}}{\Rot(-\hatx)\psiinfa} \\
 + \matrixelem*{\Rot(-\hatz)\psiinfa}{\hat{\xi}}{\Rot(-\hatz)\psiinfa}
 \bigg] + \higherorder \,.
\end{multline}
For the P-wave state we have chosen in Eq.~\eqref{eq:psi-P-asympy} it holds
that
\begin{equation}
 \Rot(-\vecn)\ket{\psiinfa} = {-}\Rot(\vecn)\ket{\psiinfa} \,.
\end{equation}
Moreover, $\Rot(\hatz)$ is simply the identity operator.
Using this to simplify the expression, we get
\begin{multline}
 \Delta\!\expval{r^2}_1^{T_1^-}\!\!\!(L) = \frac{3\abs{\gamma}^2\alpha}{\mu L}
 \ee^{-\kappa L}
 +2\Rp\bigg[
 \matrixelem*{\psiinfa}{\hat{\xi}}{\psiinfa}
 + \matrixelem*{\Rot(\hatx)\psiinfa}{\hat{\xi}}{\Rot(\hatx)\psiinfa} \\
 \null + \matrixelem*{\Rot(\haty)\psiinfa}{\hat{\xi}}{\Rot(\haty)\psiinfa}
 \bigg] + \higherorder \,.
\end{multline}
Writing out the integrand that appears in the evaluation of the matrix
elements leads to
\begin{multline}
 \psiinfa^*(\vecr)\hat{\xi}\psiinfa(\vecr)
 + \left[\Rot(\hatx)\psiinfa(\vecr)\right]^*\hat{\xi}
   \left[\Rot(\hatx)\psiinfa(\vecr)\right]
 + \left[\Rot(\haty)\psiinfa(\vecr)\right]^*\hat{\xi}
   \left[\Rot(\haty)\psiinfa(\vecr)\right] \\
 = \chi_{P\cap\setasymptotic}(\vecr)
 \frac{3 \abs{\gamma}^2 (\kappa  r+1)
 \ee^{-\kappa (2 r+u)}}{64 \pi  \kappa ^2 L^3 r^6 u^2} \Bigg\{
 4 L^2 r^2 \abs{1-\frac{\left(L^2+r^2-u^2\right)^2}{4 L^2 r^2}}
 \Big(L u^2 (\kappa  r+1) (u-r) (r+u) e^{\kappa  u} \\
 + 2 r^4 \ee^{\kappa  r}
 \left(u^2-4\expval{r^2_\infty}\right) (\kappa  u+1)\Big)
 + \left(L^2+r^2-u^2\right) \Big(L u^2 (\kappa  r+1) (u-r) (r+u) e^{\kappa u}
 \left(L^2+r^2-u^2\right) \\
 + 2 r^4 e^{\kappa  r}
 \left(u^2-4 \expval{r^2_\infty}\right) (\kappa  u+1)
 \left(r^2-u^2-L^2\right)\Big)\Bigg\} \,,
\end{multline}
noting that both $u$ and $r$ will be integrated over as described in the main
text.
We now perform the integrals in Eq.~\eqref{eq:matrix-element-integrals} with
this integrand and drop higher order terms:
\begin{multline}
 \matrixelem*{\psiinfa}{\hat{\xi}}{\psiinfa}
 + \matrixelem*{\Rot(\hatx)\psiinfa}{\hat{\xi}}{\Rot(\hatx)\psiinfa}
 + \matrixelem*{\Rot(\haty)\psiinfa}{\hat{\xi}}{\Rot(\haty)\psiinfa} \\
 = \frac{3}{32 \kappa^2}\abs{\gamma}^2 L\left(8-\kappa^2 L^2\right)
 \text{Ei}(-\kappa L) + \abs{\gamma}^2\ee^{-\kappa L}\Bigg[
 {-}\frac{L^2}{8\kappa}
 + \frac{3\left(5+8\kappa^2 \expval{r^2_\infty}\right)}{16\kappa^3}  \\
 +\frac{\ee^{-2\kappa R}}{16\kappa^4 L R} \bigg(
 \ee^{2\kappa R}\big(4\kappa^3 R^4-12 R^2 \left(\kappa+4\kappa^3
 \expval{r^2_\infty}\right)+3 R \left(8\kappa^2 \expval{r^2_\infty}+5\right) \\
 - 24\kappa \expval{r^2_\infty}\big)
 - 3\left(2\kappa^2 R^3+6 \kappa R^2+R \left(5-8\kappa^2
 \expval{r^2_\infty}\right)-16 \kappa \expval{r^2_\infty}\right)\bigg)
 \Bigg] + \higherorder \,.
\end{multline}
Putting everything back together, we get
\begin{multline}
 \Delta\!\expval{r^2}_1^{T_1^-}(L) =
 \frac{3}{16 \kappa^2}\abs{\gamma}^2 L\left(8-\kappa^2 L^2\right)
 \text{Ei}(-\kappa L)
 + \abs{\gamma}^2\ee^{-\kappa L}\bigg[-\frac{L^2}{4\kappa}
 + \frac{3\left(5+8\kappa^2 \expval{r^2_\infty}\right)}{8\kappa^3} \\
 + \frac{\ee^{-2\kappa R}}{8\kappa^4 L R} \bigg(
 \ee^{2\kappa R}\big(4\kappa^3 R^4-12 R^2 \left(\kappa+4\kappa^3
 \expval{r^2_\infty}\right)
 + 3 R \left(8\kappa^2 \expval{r^2_\infty}+5\right) \\
 - 48\kappa  \expval{r^2_\infty}\big)
 -3\left(2 \kappa^2 R^3+6 \kappa  R^2+R \left(5-8\kappa^2
 \expval{r^2_\infty}\right)-16 \kappa  \expval{r^2_\infty}\right)\bigg)
 + \frac{3\alpha}{\mu L}\Bigg] + \higherorder \,.
\end{multline}
As for $A_1^+$ S-wave sates, we can absorb the constants that appear in this expression into a single constant $a$ that needs to be fitted.
Doing that, we arrive at Eq.~\eqref{eq:radius-shift-T1m} in the main text.
\end{widetext}

\bibliographystyle{apsrev4-1}

\end{document}